\documentclass[journal]{IEEEtran}
\usepackage[cmex10]{amsmath}
\usepackage{amssymb,amsthm,amsfonts,mathrsfs,xparse}
\usepackage{graphicx}
\usepackage{dsfont,enumitem,stackrel}
\usepackage{color}
\usepackage{mathtools,amssymb,bm,mathabx}
\usepackage{amstext}
\usepackage{array}
\usepackage{algorithmicx}
\usepackage[ruled]{algorithm}
\usepackage{algpseudocode}
\usepackage{algpascal}
\usepackage{algc}
\usepackage{cases}
\usepackage{pgfplots}
\usepackage{accents}
\usepackage{caption}
\usepackage{float}
\usepackage{cite}
\usepackage [autostyle, english = american]{csquotes}
\usepackage{hyperref}
\theoremstyle{remark}

\newcommand\ASTART{\bigskip\noindent\begin{minipage}[b]{0.5\linewidth}}
	
	\newcommand\AENDSKIP{\end{minipage}\bigskip}
\newcommand\AEND{\end{minipage}}
\ifCLASSOPTIONcaptionsoff
\usepackage[nomarkers]{endfloat}
\let\MYoriglatexcaption\caption
\renewcommand{\caption}[2][\relax]{\MYoriglatexcaption[#2]{#2}}
\fi
\theoremstyle{plain}
\newtheorem{thm}{\textbf{Theorem}}
\newtheorem{lem}{\textbf{Lemma}}
\newtheorem{prop}{\textbf{Proposition}}

\theoremstyle{definition}
\newtheorem{defn}{\textbf{Definition}}

\theoremstyle{remark}
\newtheorem{rem}{\bf Remark}

\newcommand*{\rom}[1]{\expandafter\@slowromancap\romannumeral #1@}

\graphicspath{{./Figures/}} 
\begin{document}
%
%\onecolumn
\title{Distribution-aware $\ell_1$ Analysis Minimization}
\author{Raziyeh Takbiri, Sajad~Daei
\thanks{R.Takbiri is with the school of Electrical Engineering, Iran University of Science \& Technology, Iran. S.Daei is with the Information Science and Engineering Department, KTH University, Sweden}}
\maketitle

\begin{abstract}
This work is about recovering an analysis-sparse vector, i.e. sparse vector in some transform domain, from under-sampled measurements. In real-world applications, there often exist random analysis-sparse vectors whose distribution in the analysis domain are known. To exploit this information, a weighted $\ell_1$ analysis minimization is often considered. The task of choosing the weights in this case is however challenging and non-trivial. In this work, we provide an analytical method to choose the suitable weights. Specifically, we first obtain a tight upper-bound expression for the expected number of required measurements. This bound depends on two critical parameters: support distribution and expected sign of the analysis domain which are both accessible in advance. Then, we calculate the near-optimal weights by minimizing this expression with respect to the weights. Our strategy works for both noiseless and noisy settings. Numerical results demonstrate the superiority of our proposed method. Specifically, the weighted $\ell_1$ analysis minimization with our near-optimal weighting design considerably needs fewer measurements than its regular $\ell_1$ analysis counterpart.
\end{abstract}


% \fi
\IEEEpeerreviewmaketitle

\section{Introduction}
 \IEEEPARstart{C}{ompressed} sensing (CS) is a well-established theory during the last decade providing a new framework to recover a sparse vector from far few measurements. To achieve successful recovery, these measurements must satisfy a restricted isometry property (RIP) \cite{candes2008restricted}. The traditional CS, initiated by the works \cite{candes2005decoding,donoho2005sparse,donoho2006high}, amounts to reconstructing signals that are sparse in an orthonormal basis. However, in practical scenarios, there often exist examples in which the sparsity is not expressed in terms of some orthonormal bases but in terms of a few atoms of a redundant and coherent dictionary. This means that the signal of interest e.g. $\bm{x}\in\mathbb{R}^n$ can be expressed as $\bm{x}=\bm{D}_{n\times N}\bm{\alpha}_{N\times 1}$ where $n\ll N$. There exist two approaches in the literature to deal with such signals: $\ell_1$ synthesis and $\ell_1$ analysis; the former seeks for the coefficient vector $\bm{\alpha}$ and considers the problem
  \begin{align}
  &\min_{\bm{z}\in\mathbb{R}^N}\|\bm{z}\|_1:=\sum_{i=1}^{N}|z_i|\nonumber\\
  &\bm{y}_{m\times 1}=\bm{A}\bm{D}\bm{z},
  \end{align}
 while the latter tries to estimate the original signal $\bm{x}\in\mathbb{R}^n$ by solving
 \begin{align}\label{eq.l1analysis}
 \mathsf{P}_{\bm{\Omega}}:~~&\min_{\bm{z}\in\mathbb{R}^n}\|\bm{\Omega z}\|_1:=\sum_{i=1}^{N}|(\bm{\Omega z})_i|\nonumber\\
 &\bm{y}_{m\times 1}=\bm{A}\bm{z},
 \end{align}
 to estimate first the coefficients $\bm{\alpha}\in\mathbb{C}^{N}$ and then the interested signal $\bm{x}=\bm{D}\bm{\alpha}$
 where $\bm{\Omega}:=[\bm{\omega}_1,..., \bm{\omega}_p]^T\in\mathbb{R}^{p\times n}, p\gg n$ is a redundant matrix called analysis operator. In simple words, $\ell_1$ synthesis approach operates in the coefficient domain and then pulls back the corresponding estimates to the signal domain in $\mathbb{R}^n$ while $\ell_1$ analysis tries to examine/analyze the features of $\bm{x}$ in the analysis domain using the projections $\bm{\Omega x}=[\langle \bm{\omega}_1,\bm{x}\rangle, ..., \langle \bm{\omega}_p,\bm{x}\rangle]^T$. Although $\ell_1$ synthesis formulation has been more studied in the context of CS, it has major drawbacks. In a nutshell, besides the large space of optimization search ($n\ll N$), one can not hope to reach an accurate solution for the original signal by $\ell_1$ synthesis approach \cite{candes2011compressed,genzel2017ell}. In fact, the measurement matrix $\bm{A D}$ may no longer satisfy the required conditions of conventional CS such as RIP.  
 
 Although $\ell_1$ analysis has gained growing attention during the recent years \cite{daei2019error,genzel2017ell,daei2019living}, most of the literature so far on $\ell_1$ analysis minimization has focused on the case where the only prior information about the signal of interst is its sparsity in a domain, i.e. $\bm{\Omega x}$ is sparse for some analysis operator $\bm{\Omega}$. However, in many real-world applications, $\bm{x}$ is a random vector and the distribution of $\bm{\Omega x}$ is known in advance. For example, if $\bm{x}$ is an image and $\bm{\Omega}$ is a wavelet or discrete cosine transform (DCT) matrix, then the first elements of $\bm{\Omega x}$ are more likely to be non-zero (i.e. to be in the support set) than others. In such cases, one is interested in modifying $\mathsf{P}_{\bm{\Omega}}$ in order to incorporate such extra distributional/statistical information into the recovery procedure. This new setting is considered in the following optimization problem:
 \begin{align}\label{eq.weightedl1}
 \mathsf{P}_{\bm{\Omega},\bm{v}}:~~&\min_{\bm{z}\in\mathbb{R}^n}\|\bm{\Omega z}\|_{1,\bm{v}}:=\sum_{i=1}^{p}v_i|(\bm{\Omega z})_i|\nonumber\\
 &\bm{y}_{m\times 1}=\bm{A}\bm{z},
 \end{align}
 where $\bm{v}=\{v_1,\dots,v_p\}$ are some positive scalars served to penalize the directions that are less likely to be on the analysis support i.e. locations/indices that $(\bm{\Omega x})_i\neq 0, i=1,...,p$.

The problem $\mathsf{P}_{\Omega,\bm{v}}$ was previously investigated in \cite{candes2011compressed}. The authors, based on numerical simulations, showed that $\mathsf{P}_{\bm{\Omega},\bm{v}}$ considerably outperforms $\mathsf{P}_{\bm{\Omega}}$ when the weights $\bm{v}$ are chosen in a re-weighted manner as
\begin{align}
v_i=\frac{1}{|(\bm{\Omega} \widehat{\bm{ x}})_i|+\epsilon},
\end{align}
 where $\widehat{\bm{x}}$ is the solution of the previous iteration. However, the authors left it unanswered how to choose the weights in an optimal and theoretical way which is a challenging task and nontrivial. Our work provides a complete answer to this question. More explicitly, we propose to choose the weights that minimize the expected number of required measurements in $\mathsf{P}_{\Omega,\bm{v}}$. To do so, we will use the recent result of \cite{genzel2017ell} which obtains a bound describing the required sample complexity of $\mathsf{P}_{\Omega}$.
 
 \subsection{Contributions, Prior Arts and Key Differences}
 As our strategy to find the weights in $\mathsf{P}_{\Omega,\bm{v}}$ is highly dependent on sample complexity computations, we review the recent results providing expressions for the required number of measurements in $\mathsf{P}_{\Omega}$.
 
 Sparse recovery is a more well-known concept than recovery of analysis-sparse signals. It has been up to now proved that an $s$-sparse vector $\bm{x}\in\mathbb{R}^n$ can be recovered using sub-Gaussian measurements provided that the sample complexity is of order $\mathcal{O}(s\log(\frac{2s}{n}))$ \cite{candes2006robust,amelunxen2013living}. However, from the recent results of \cite{genzel2017ell,daei2019living,daei2019error,daei2018sample}, it has been turned out that sparsity per se does not explain the required sample complexity of $\mathsf{P}_{\Omega}$ or $\mathsf{P}_{\Omega,\bm{v}}$. The more recent work of \cite{genzel2017ell} has introduced an upper-bound describing the required sample complexity of $\mathsf{P}_{\Omega}$. This bound explicitly depends on the row correlations of $\bm{\Omega}$ in the support and off-support locations i.e. $\langle \bm{\omega_i},\bm{\omega_j}\rangle, \forall i,j\in\mathcal{S}$ and $\langle \bm{\omega_i},\bm{\omega_j}\rangle, \forall i,j\in\overline{\mathcal{S}}$, respectively. This bound is numerically observed to be tight for different classes of analysis operators. Due to the dependence on the unknown signal characteristics namely $\mathcal{S}$ and $\overline{\mathcal{S}}$, the derivation of this bound is impossible in advance.

 Apart from this, exploiting additional information has a long history in the literature of CS e.g. see \cite{daei2019exploiting,daei2018improved,daei2019distribution,flinth2016optimal,diaz2017compressed}. In \cite{flinth2016optimal}, the authors proposed an approach for using deterministic prior information in enhancing the sparse recovery performance. The prior information was in the form of a few sets contributing to the support with a given accuracy. The strategy was to propose a weighted $\ell_1$ minimization and choose the weights that minimize the required number of measurements which was obtained in \cite{amelunxen2013living} for simple sparse signals.
 
 In \cite{diaz2017compressed}, the authors proposed a strategy to incorporate statistical information about the interested sparse signal. They considered a distribution for the support of the sparse signal and proposed a weighted $\ell_1$ minimization. The weights are such chosen to minimize an expression describing the expected number of measurements required for exact recovery.
 
In \cite{daei2019exploiting,daei2019distribution}, the authors incorporated prior deterministic and statistical block information into the block sparse recovery via penalizing the weights assigned to each block via weighted $\ell_{1,2}$ minimization.

In \cite{daei2018improved}, the authors exploited deterministic prior information in order to minimize the required number of measurements in weighted $\ell_1$ analysis minimization. They proposed an upper-bound for the number of measurements one needs for exact recovery. Their bound is not tight i.e. it does not explain the required number of measurements for highly coherent and redundant analysis operators. This is due to the fact that the bound has direct relation to the condition number of analysis operator which has high values for coherent and redundant operators. Consequently, their obtained weights which are found by minimizing this imprecise bound leads to inaccurate weights for highly coherent analysis operators. 

Unlike \cite{daei2018improved}, our proposed near-optimal weights in this paper are obtained by minimizing a novel tight bound. This bound specifies the required sample complexity of $\mathsf{P}_{\Omega,\bm{v}}$. Our proposed bound is suitable for many cases of analysis operators including highly coherent and redundant. To achieve this threshold bound for $\mathsf{P}_{\Omega,\bm{v}}$, we use the approach of \cite{genzel2017ell}. While deriving the proposed bound of \cite{genzel2017ell} is impossible for $\mathsf{P}_{\Omega}$ due to the reasons mentioned above, our proposed bound for the expected sample complexity of $\mathsf{P}_{\Omega,\bm{v}}$ is achievable and tractable. Specifically, our proposed bound describing the expected sample complexity of $\mathsf{P}_{\Omega,\bm{v}}$ does explicitly depend on two main parameters: support distribution, i.e. $\mathbb{P}(k\in\mathcal{S})$, and expected sign mean of the coefficients in the analysis domain, i.e. $\mathds{E}({\rm sgn}(\bm{\Omega x}))$ where ${\rm sgn}(\cdot)$ returns the component-wise sign of a vector. Both of these parameters are accessible in advance and depend on the statistical information (or distribution) of the interested signal. By having this information in advance, we are able to calculate the near-optimal weights which indeed minimize the expected required number of measurements. Simulation results verify the effectiveness and superiority of our proposed strategy over regular $\ell_1$ analysis optimization in terms of reconstruction error. 
\subsection{Outline}
In Section \ref{section.convexgeometry}, we review some of the basic concepts in convex geometry which are required to obtain the required  number of measurements. Section \ref{sec.main_result} amounts to obtaining an upper-bound for the required number of measurements as well as providing a novel strategy to find the near-optimal weights. Section \ref{sec.simulation} is dedicated to synthetic and real-world experiments to evaluate the performance of our proposed method. Finally, the paper is concluded in Section \ref{sec.conclusion}.
\subsection{Notation}
Vectors e.g. $\bm{x}\in\mathbb{R}^n$ and matrices e.g. $\bm{X}\in\mathbb{R}^{n_1\times n_2}$ are denoted by small and large bold-face letters, respectively. The support of a sparse vector $\bm{x}\in\mathbb{R}^n$ is defined as the set of locations with non-zero values i.e. $\mathcal{S}:=\{i\in\{1,..., n\}:x_i\neq 0\}$. Also, the support complement is denoted by $\overline{\mathcal{S}}:=\{1,..., n\}\setminus \mathcal{S}$. $\bm{X}_{\mathcal{S}}\in\mathbb{R}^{|\mathcal{S}|\times n}$ refers to the matrix $\bm{X}\in\mathbb{R}^{m\times n}$ restricted to the rows indexed by $\mathcal{S}$. Also, $\bm{x}_{\mathcal{S}}\in\mathbb{R}^{|\mathcal{S}|}$ is a restriction of $\bm{x}\in\mathbb{R}^n$ to the indices $\mathcal{S}$. $\bm{I}_n$ is the identity matrix of size $n\times n$.   $\bm{x}\odot\bm{y}$ is used for element-wise multiplication of two equal-length vectors $\bm{x}$ and $\bm{y}$. We show a standard Gaussian random vector with i.i.d. elements by $\bm{g}\in\mathbb{R}^n$ i.e. $\bm{g}~\mathcal{N}(\bm{0},\bm{I}_n)$. The error function is defined as ${\rm erf}(x):=\frac{2}{\sqrt{\pi}}\int_{0}^x {\rm e}^{-t^2}dt$. The $\ell_1$ and $\ell_2$ and $\ell_{\infty}$ norms are respectively defined by $\|\bm{x}\|_1:=\sum_{i=1}^{n}|x_i|$ and $\|\bm{x}\|_2=\sqrt{\sum_{i=1}^{n}|x_i|^2}$ and $\|\bm{x}\|_{\infty}:=\max_i|x_i|$. $1_{i\in\mathcal{E}}$ is the indicator function which is defined as $1_{i\in\mathcal{E}}=\left\{\begin{array}{cc}
1& i\in \mathcal{E}\\
0& o.w.
\end{array}\right\}$. $a\wedge b$ means $\min\{a,b\}$.
Component-wise sign of a vector $\bm{x}$ is defined as ${\rm sgn}(\bm{x}):=[{\rm sgn}(x_1),..., {\rm sgn}(x_n)]^T$. For a set $C$, we show the polar of that set by $C^{\circ}$. $\mathcal{P}_\mathcal{C}(\bm{x})$ is the projection of $\bm{x}\in \mathbb{R}^n$ onto the set $\mathcal{C}$ and is defined as $\mathcal{P}_\mathcal{C}(\bm{x})=\underset{\bm{z} \in \mathcal{C}}{\arg\min}~\|\bm{z}-\bm{x}\|_2$. ${\rm null}(\bm{A})$ is used to denote the null space of the matrix $\bm{A}\in\mathbb{R}^{m\times n}$ which is defined as $\{\bm{z}\in\mathbb{R}^n:\bm{A z}=\bm{0}\}$.
 \section{Convex Geometry}\label{section.convexgeometry}
 In this section, a review of basic concepts of convex geometry is provided.
 \subsection{Descent Cones}
 Consider a fixed point $\bm{x}\in\mathbb{R}^n$ that has a simple structure e.g. sparsity, low-rankness, analysis-sparsity. Let $f:\mathbb{R}^n\rightarrow \mathbb{R}$ be a convex function that reflects the specific structure of $\bm{x}$.
The set of directions $\bm{z}\in\mathbb{R}^n$ that lead to a reduction in the value of $f$ form a convex cone called descent cone which is defined as below:
 \begin{align}\label{eq.descent cone}
 \mathcal{D}(f,\bm{x})=\bigcup_{t\ge0}\{\bm{z}\in\mathbb{R}^n: f(\bm{x}+t\bm{z})\le f(\bm{x})\}.
 \end{align}
 The descent cone of the convex function $f$ at point $\bm{x}$ is a convex set. Descent cone of convex functions is linked to the subdifferential (c.f. \cite[Chapter 23]{rockafellar2015convex}) provided by:
 \begin{align}\label{eq.D(f,x)}
 \mathcal{D}^{\circ}(f,\bm{x})=\mathrm{cone}(\partial f(\bm{x})):=\bigcup_{t\ge0}t\partial f(\bm{x}).
 \end{align}
 \subsection{Statistical Dimension and Optimality Condition}
 To extract the structure from a few linear measurements $\bm{y}=\bm{A x}$, one may solve the following convex optimization:
 \begin{align}\label{prob.general}
 &\mathsf{P}_f: \min_{\bm{z}\in\mathbb{R}^n}f(\bm{z})\nonumber\\
 &s.t. ~~\bm{y}_{m\times 1}=\bm{A z}.
 \end{align} 
 The boundary of failure and success in linear inverse problems, i.e. the minimum $m$ that is needed for \eqref{prob.general} to reach exact recovery with a probability of $.5$, is described by the statistical dimension of the generated descent cone which is defined as below.
 \begin{defn}{Statistical Dimension}\cite{amelunxen2013living}:
 	Consider the closed convex cone $\mathcal{C}\subseteq\mathbb{R}^n$. The statistical dimension of $\mathcal{C}$ is defined as:
 	\begin{align}\label{eq.statisticaldimension}
 	\delta(\mathcal{C}):=\mathds{E}\|\mathcal{P}_\mathcal{C}(\bm{g})\|_2^2=\mathds{E}~\mathrm{dist}^2(\bm{g},\mathcal{C}^\circ).
 	\end{align}
 \end{defn}
 The statistical dimension extends the concept of linear subspaces to convex cones. Furthermore, $\delta(\mathcal{D}(f,\bm{x}))$ determines the precise location of transition from failure to success in $\mathsf{P}_f$. Exact calculation of the statistical dimension is highly challenging for different structures. For analysis-sparsity and gradient sparsity, it has been proved in \cite{daei2019error} that the following bound explains $\delta(\mathcal{D}(f,\bm{x}))$ well:
 \begin{align}
 B_u=\inf_{t\ge 0}\mathds{E}_{\bm{g}}\inf_{z\in\partial f(\bm{x})}\|\bm{g}-t\bm{z}\|_2^2.
 \end{align}
 
 In the following, we provide conditions in which $\mathsf{P}_f$ succeeds in the noise-free case.
 \begin{prop}\cite[Proposition 2.1]{chandrasekaran2012convex} Optimality condition: Let $f$ be a proper convex function. The vector $\bm{x}\in \mathbb{R}^n$ is the unique optimal point of $\mathsf{P}_f$ if and only if $\mathcal{D}(f,\bm{x})\cap \mathrm{null}(\bm{A})=\{\bm{0}\}$.
 \end{prop}
 The next theorem determines the number of measurements needed for successful recovery of $\mathsf{P}_f$ for any proper convex function $f$.
 \begin{thm}\label{thm.Pfmeasurement}\cite[Theorem 2]{amelunxen2013living}:
 	Let $f:\mathbb{R}^n\rightarrow \mathbb{R}\cup \{\pm\infty\}$ be a proper convex function and $\bm{x}\in \mathbb{R}^n$ a fixed vector. Suppose that $m$ independent Gaussian linear measurements of $\bm{x}$ are observed via $\bm{y}=\bm{Ax} \in \mathbb{R}^m$. If
 	\begin{align}
 	m\ge \delta(\mathcal{D}(f,\bm{x}))+\sqrt{8\log(\frac{4}{\eta})n},
 	\end{align}
 	for a given probability of failure (tolerance) $\eta \in [0,1]$, then, we have
 	\begin{align}
 	\mathds{P}(\mathcal{D}(f,\bm{x})\cap \mathrm{null}(\bm{A})=\{\bm{0}\})\ge 1-\eta.
 	\end{align}
 	Moreover, if
 	\begin{align}
 	m\le \delta(\mathcal{D}(f,\bm{x}))-\sqrt{8\log(\frac{4}{\eta})n},
 	\end{align}
 	then,
 	\begin{align}
 	\mathds{P}(\mathcal{D}(f,\bm{x})\cap \mathrm{null}(\bm{A})=\{\bm{0}\})\le \eta.
 	\end{align}
 \end{thm}
 For random vectors, the exact recovery arguments is modified and the probability of success does depend on both probability density function (pdf) of $\bm{x}$ and $\bm{A}$. In fact, there is a phase transition phenomenon analogous to Theorem \ref{thm.Pfmeasurement}.
\begin{thm}\cite[Theorem 2.4]{diaz2017compressed}. Let $\bm{x}\in\mathbb{R}^n$ be a random vector and $\bm{A}\in\mathbb{R}^{m\times n}$ is a random measurement matrix. For a given weight vector $\bm{v}$ and a tolerance probability $\eta$, we have:
	\begin{align}
	&\mathds{P}\{	m\ge \delta(\mathcal{D}(f,\bm{x}))+\sqrt{8\log(\frac{4}{\eta})n}\}\ge 1-\frac{\eta}{2}\rightarrow\nonumber\\
	 &{\rm prob}(\bm{v})\ge 1-\eta\\
	&\mathds{P}\{m\le \delta(\mathcal{D}(f,\bm{x}))-\sqrt{8\log(\frac{4}{\eta})n}\}\ge 1-\frac{\eta}{2}\rightarrow\nonumber\\
	&{\rm prob}(\bm{v})\le \eta ,
	\end{align} 
	where ${\rm prob}(\bm{v}):=\mathds{P}_{\bm{A},\bm{x}}\{ \mathsf{P}_{\bm{\Omega},\bm{v}} ~\text{is successful for given} ~\bm{x} \text{ and} ~\bm{A}\}$.
\end{thm}
The above theorem suggests that the more the pdf of $\delta(\mathcal{D}(f,\bm{x}))$ is concentrated around its mean, the phase transition becomes narrower. Since $\delta(\mathcal{D}(f,\bm{x}))$ is a random variable, this work amounts to optimizing a deterministic expression.
Therefore, we introduce the expected statistical dimension in the following:
\begin{defn}\cite[Definition 2.5] {diaz2017compressed}(Expected statistical dimension): Let $\bm{x}\in\mathbb{R}^n$ be a random vector and $\bm{A}\in\mathbb{R}^{m\times n}$ is a random measurement matrix. For a given vector of weights $\bm{v}$, the expected statistical dimension is defined as
	\begin{align}
	\overline{\delta}(\bm{v}):=\mathds{E}_{x}[\delta(\mathcal{D}(f,\bm{x}))].
	\end{align}
\end{defn}
 \section{Main Results}\label{sec.main_result}
 In this section, we first obtain an upper-bound for the required sample complexity i.e. $\delta(\mathcal{D}(\|\Omega\cdot\|_{1,\bm{v}},\bm{x}))$. Then, we provide our strategy to obtain near-optimal weights.
 \subsection{Expected Sample Complexity}
 We begin with a lemma that allows us to bound ${\delta(\mathcal{D}(\|\bm{\Omega}\cdot\|_{1,\bm{v}},\bm{x}))}$ which roughly describes the required number of measurements that $\mathsf{P}_{\bm{\Omega},\bm{v}}$ needs for exact recovery.
 \begin{lem}\label{lem.upper1}
 For any $\bm{x}\in\mathbb{R}^n$ and any positive weights $\bm{v}\in\mathbb{R}^p$, the statistical dimension of the descent cone $\mathcal{D}(\|\bm{\Omega} \cdot\|_{1,\bm{v}},\bm{x})$ is bounded from above by the following expression:
\begin{align}\label{eq.upper1}
&\delta(\mathcal{D}(\|\bm{\Omega}\cdot\|_{1,\bm{v}},\bm{x}))\le \inf_{t> 0}\inf_{\lambda> 0}\Bigg\{n+\lambda^2\sum_{i,j\in\mathcal{S}}(tv_i)(tv_j)\bm{\omega}_i^T\bm{\omega}_j\nonumber\\
&{\rm sgn}(\bm{\omega}_i^T\bm{x}){\rm sgn}(\bm{\omega}_j^T\bm{x})-2\sum_{i\in\overline{\mathcal{S}}}\lambda \|\bm{\omega}_i\|_2 {\rm erf}(\frac{tv_i}{\sqrt{2}})+\nonumber\\
&\lambda^2\sum_{i,j\in\overline{\mathcal{S}}}\frac{(\bm{\omega}_i^T\bm{\omega}_j)^2}{\|\bm{\omega}_i\|_2\|\bm{\omega}_j\|_2}\bigg[{\rm erf}(\frac{tv_{\min}}{\sqrt{2}})-h(tv_{\max})(tv_i)(tv_j)\bigg]\Bigg\}.
\end{align}
 \end{lem}
Proof. See Appendix \ref{proof.lemma_upper}.
 \begin{rem}(Special case)
 When there are equal weights i.e. $\bm{v}(1)=\bm{v}(2)=...=\bm{v}(p)=1$, \eqref{eq.upper1} reduces to the expression
 \begin{align}\label{eq.upper_genzel}
&\delta(\mathcal{D}(\|\bm{\Omega}\cdot\|_{1,\bm{v}},\bm{x}))\le \nonumber\\
&\inf_{t> 0}\inf_{\lambda>0}\{n+t^2\lambda^2\Big(\|\bm{\Omega}_{\mathcal{S}}^T{\rm sgn}(\bm{\Omega x})_{\mathcal{S}}\|_2^2)-h(t)\sum_{i,j\in\overline{\mathcal{S}}}\frac{(\bm{\omega}_i^T\bm{\omega}_j)^2}{\|\bm{\omega}_i\|_2\|\bm{\omega}_j\|_2}\Big)\nonumber\\
&+\Big(\lambda^2\sum_{i,j\in\overline{\mathcal{S}}}\frac{(\bm{\omega}_i^T\bm{\omega}_j)^2}{\|\bm{\omega}_i\|_2\|\bm{\omega}_j\|_2}-2\lambda \sum_{i\in\overline{\mathcal{S}}} \|\bm{\omega}_i\|_2 \Big){\rm erf}(\frac{t}{\sqrt{2}})\}
 \end{align}
 which matches the sample complexity obtained in \cite[Proposition 6.13, Equation 6.16]{genzel2017ell}.
 \end{rem}
 \begin{figure}[t]
 	\centering
 	\includegraphics[scale=.5]{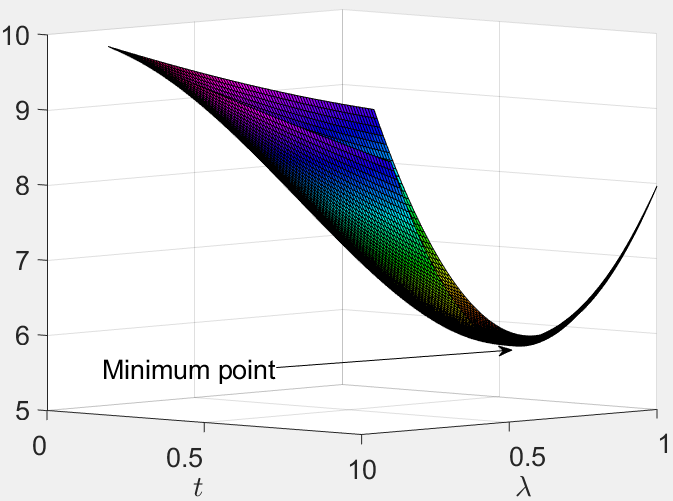}
 	\caption{This figure shows the expression obtained in Lemma \ref{lem.upper1} for typical examples of weights and $\bm{x}$. As it turns out, there is a point $(t,\lambda)$ in which the upper-bound expression is minimized.}\label{fig.lem1}
 \end{figure}
According to the above lemma, the upper-bound expression \eqref{eq.upper1} depends on the characteristics of the desired signal $\bm{x}$ such as  ${\rm sgn}(\bm{\Omega}\bm{x})$, and $\mathcal{S}$ and $\overline{\mathcal{S}}$ which are all unknown in advance. Fortunately, in the following theorem, we obtain a bound for the expected statistical dimension that is accessible/tractable.
 \begin{thm}\label{thm.main}
 	Let $f_{{d}_k}(\cdot)$ be the pdf of $d_k:={(\Omega x)_k}$ for $k=1,\dots p$. Then, for any $\bm{v}\in\mathbb{R}_{+}^p$, the expected number of required measurements for $\mathsf{P}_{\bm{\Omega},\bm{v}}$ which is denoted by $\overline{m}_{\bm{v}}$ satisfies 
 	\begin{align}\label{eq.upper_expected}
 	&\mathds{E}_{\bm{x}}[\delta(\mathcal{D}(\|\bm{\Omega} \cdot\|_{1,\bm{v}},\bm{x}))]\le\nonumber\\
 	& \inf_{t\ge 0}\Big\{ n-\frac{\Big(\sum_{k=1}^p(1-\beta_k)\|\bm{\omega}_k\|_2{\rm erf}(\frac{tv_k}{\sqrt{2}})\Big)^2}{F(t,\bm{v})}\Big\},\nonumber\\
 	&F(t,\bm{v})=\sum_{k,k'=1}^p\Bigg[(tv_k)(tv_{k'})\bm{\omega_k}^T\bm{\omega}_k\sigma_k\sigma_{k'}\beta_k\beta_{k'}+\frac{(\bm{\omega_k}^T\bm{\omega}_k)^2}{\|\bm{\omega}_k\|_2\|\bm{\omega}_{k'}\|_2}\nonumber\\
 	&\Big({\rm erf}(\frac{tv_{\rm min}}{\sqrt{2}})-h(tv_{\rm max})(tv_k)(tv_{k'})\Big)(1-\beta_k)(1-\beta_{k'})\Bigg],
 	\end{align}
 	where
 	\begin{align}
 	&h(t):=\sqrt{\frac{2}{\pi}}\frac{{\rm e}^{-\frac{t^2}{2}}}{t}+{\rm erf}(\frac{t}{\sqrt{2}})-1,\nonumber\\
 	&\beta_k:=\mathds{P}\{k\in\mathcal{S}\},\nonumber\\
 	&\sigma_k:=\int_{h}{\rm sgn}(h)f_{{d}_k}(h){\rm d}h 
 	\end{align}
 \end{thm}
 \begin{rem}(Effective parameters in expected statistical dimension):
 The expression \eqref{eq.upper_expected} does depend on two major parameters: 1. The probability of each analysis element to be in the support $\mathcal{S}$ denoted by $\beta_k, k=1,..., p$, namely the support distribution. 2. The expected sign values of the analysis coefficients denoted by $\sigma_k, k=1,..., p$. These two parameters are both accessible in advance. Therefore, in contrast to \eqref{eq.upper1} or \eqref{eq.upper_genzel} \cite[Equation 6.16]{genzel2017ell},  \eqref{eq.upper_expected} is achievable.
 \end{rem}

 \subsection{Near-optimal weights}
 In this section, based on the obtained bound in Theorem \ref{thm.main}, we provide an efficient strategy to find the near-optimal weights.
Our aim is to find weights that minimize the expression \eqref{eq.upper_expected}. In fact, we are seeking solutions to solve the following optimization problem:
 \begin{align}
  \bm{v}^\star=\mathop{\arg \min}_{\bm{v}>0}\inf_{t> 0}\Big\{ n-\frac{\Big(\sum_{k=1}^p(1-\beta_k)\|\bm{\omega}_k\|_2{\rm erf}(\frac{tv_k}{\sqrt{2}})\Big)^2}{F(t,\bm{v})}\Big\}
 \end{align}
 where $F(t,\bm{v})$ is provided in Theorem \ref{thm.main}. As scaling of $\bm{v}$ does not affect the optimization $\mathsf{P}_{\Omega,\bm{v}}$ in \eqref{eq.weightedl1}, we redefine $t\bm{v}$ as a single variable $\bm{v}$ and transform the latter optimization into
 \begin{align}\label{eq.cost_func}
\bm{v}^\star=\mathop{\arg \min}_{\bm{v}\in\mathbb{R}_{++}^p}\Big\{ n-\frac{\Big(\sum_{k=1}^p(1-\beta_k)\|\bm{\omega}_k\|_2{\rm erf}(\frac{v_k}{\sqrt{2}})\Big)^2}{F_1(\bm{v})}:=\overline{\delta}(\bm{v})\Big\},
 \end{align}
 where 
 \begin{align}
 &F_1(\bm{v})=\sum_{k,k'=1}^p\Bigg[v_k v_{k'}\bm{\omega}_k^T\bm{\omega}_k\sigma_k\sigma_{k'}\beta_k\beta_{k'}+\frac{(\bm{\omega}_k^T\bm{\omega}_k)^2}{\|\bm{\omega}_k\|_2\|\bm{\omega}_{k'}\|_2}\nonumber\\
 &\Big({\rm erf}(\frac{v_{\rm min}}{\sqrt{2}})-h(v_{\rm max})v_k v_{k'}\Big)(1-\beta_k)(1-\beta_{k'})\Bigg].
 \end{align}
 Numerical experiments show that the expression \eqref{eq.cost_func} is near $\mathds{E}_{\bm{x}}\delta(\mathcal{D}(\|\bm{\Omega} \cdot\|_1,\bm{x}))$. As such, since the weights that minimize $\mathds{E}_{\bm{x}}\delta(\mathcal{D}(\|\bm{\Omega} \cdot\|_1,\bm{x}))$ are unique (analogous to the simple sparse case in \cite{flinth2016optimal}), we reach to the conclusion that the near-optimal weights are uniquely determined by minimizing \eqref{eq.cost_func} given the additional constraint $\|\bm{v}\|_{\infty}=1$. For this task, we design a alternative vector optimization method in Algorithm \ref{alg.optweights} to obtain the near optimal weights. To find the vector that minimizes \eqref{eq.cost_func}, Algorithm \ref{alg.optweights} converts the vector optimization into several scalar optimization problems (with scalar variables) which then can be solved using standard scalar optimization methods (e.g. Golden Section Search (GSS) in \cite[Algorithm 2]{daei2018optimal}).
 
 It is worth mentioning that the term \lq\lq  near optimal\rq \rq refers to the fact that the bound \eqref{eq.upper_expected} is observed to be approximately tight (as is also the case for \eqref{eq.upper_genzel} in \cite{genzel2017ell}). Hence, we call our proposed weights near optimal as they are obtained by minimizing this near-tight bound. Our algorithm returns the near-optimal weights in a fast and efficient manner.
 \alglanguage{pseudocode}
 \begin{algorithm}[t]
 	\caption{}\label{alg.optweights}
 	\begin{algorithmic}[1]
 		\Require {$\bm{\beta}\in\mathbb{R}^p,\bm{\sigma}\in\mathbb{R}^p$, $\rm maxiter$, $\rm Tol$}
 		
 		\State 
 		Cost function: $\overline{\delta}(v_1,...,v_p)$ in \eqref{eq.cost_func}.
 		\State $v_i^{1}\leftarrow 1 ~~\forall i\in\{1,...,p\},$
 		\State $k\leftarrow 1,$
 		\Repeat 
 		
 		\For{$i=1~{\rm to}~p$}
 		\State Optimize the $i$th coordinate
 		\State $\phi(\zeta):=\overline{\delta}(\underbrace{v_1^{k+1},... ,v_{i-1}^{k+1}}_{\rm done} ,\underbrace{\zeta}_{\rm current},\underbrace{v_{i+1}^{k},..., v_{p}^{k}}_{\rm to~ do}),$
 		\State \begin{align}
 		v_i^{k+1}\leftarrow \mathop{\rm argmin}_{\zeta \in \mathbb{R}} \phi(\zeta)~~~~\text{use scalar optimization algorithms},\nonumber
 		\end{align}\label{step.scalar}
 		\EndFor
 		\State $k\leftarrow k+1$
 		\Until
 		{	$\|\bm{v}^{k}-\bm{v}^{k-1}\|_2 <\rm Tol$ or $|\overline{\delta}(\bm{v}^{k})-\overline{\delta}(\bm{v}^{k-1})|<\rm Tol$ or $k>{\rm maxiter}$, }
 		\State \textbf{Output} $\bm{v}^\star\leftarrow [v_1^k,...,v_p^k]^T,$
 	
 		\Statex
 	\end{algorithmic}
 \end{algorithm}
 \section{Simulations}\label{sec.simulation}
 In this section, we conduct some numerical experiments to evaluate the performance of our method. All of the experiments are performed using CVX package \cite{cvx} and MATLAB software. 

\begin{figure}[t]
 	\hspace{-1cm}
 	\includegraphics[scale=.3]{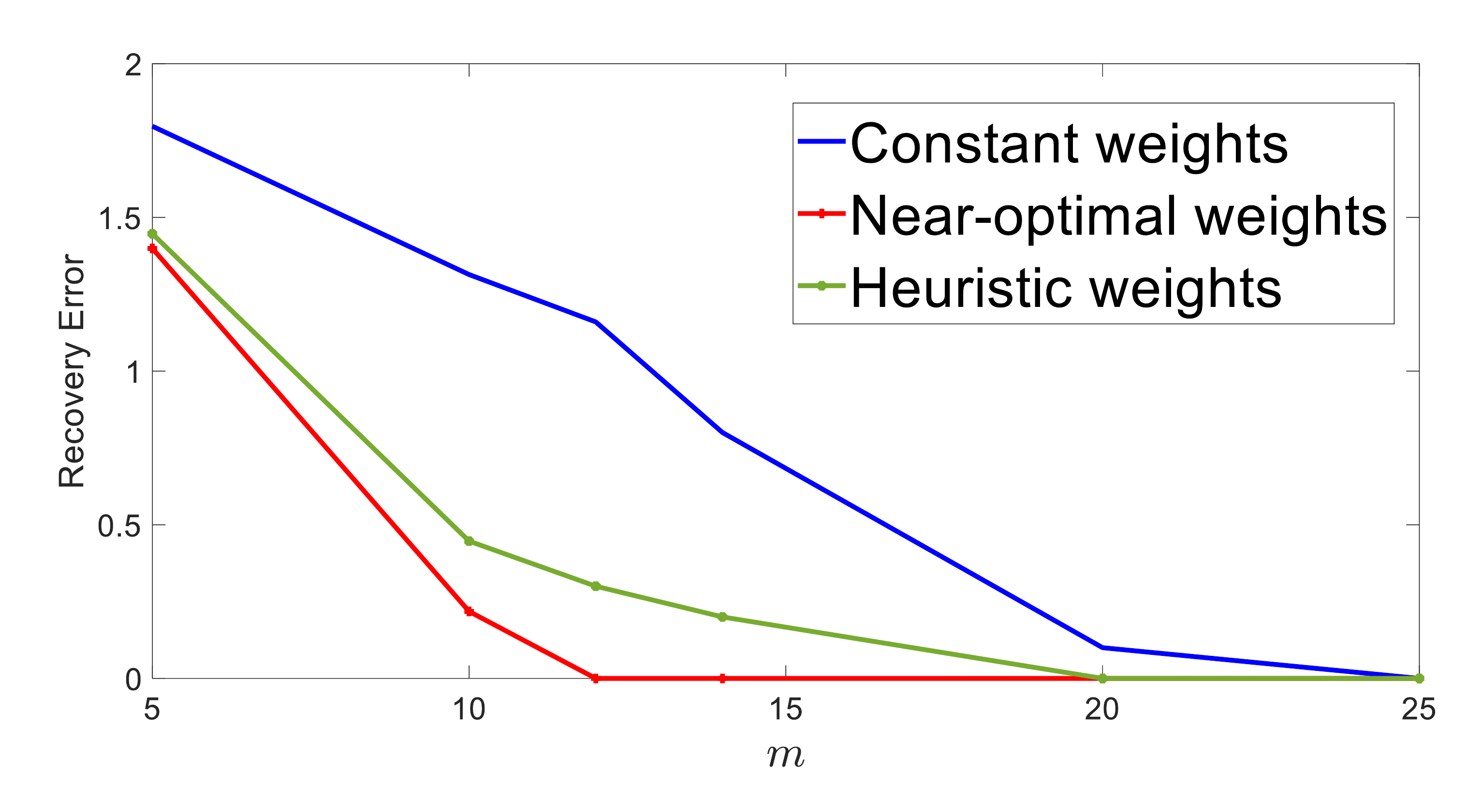}
 	\caption{This figure shows the estimated recovery error. The blue lines shows the recovery performance of $\mathsf{P}_{\bm{\Omega}}$ with constant weights. The green lines represent the performance of $\mathsf{P}_{\bm{\Omega},\bm{v}}$ with the heuristic weights $v_i=1-\beta_i, i=1..., p$  while the red line show the performance of our proposed near optimal weights obtained from Algorithm \ref{alg.optweights}}\label{fig.recovery_error}
 \end{figure}

 In the first experiment, we examine an example in which the signal of interest is random analysis-sparse with parameters $p=34, n=30$. To construct $\bm{\Omega}$, we follow the approach of \cite{kabanava2015analysis,nam2013cosparse} and first obtained orthonormal basis of the column span of a random matrix. Then, a non-tight frame $\bm{\Omega}$ is generated by changing the $\ell_2$ norms of rows. The support $\mathcal{S}$ is randomly selected from the subset $\{1,...,p\}$ with known distribution $\beta$ and the analysis signal is generated as $\bm{x}={\rm null}(\bm{\Omega}_{\overline{\mathcal{S}}})\bm{c}$ where $\bm{c}$ is a random vector uniformly distributed on the uniform sphere. We form the measurement matrix $\bm{A}_{m\times n}$ from an i.i.d. Gaussian distribution and perform the signal recovery using three methods: $\ell_1$ analysis, weighted $\ell_1$ analysis (Problem $\mathsf{P}_{\bm{\Omega},\bm{v}}$) with near optimal and heuristic weights which are provided respectively via Algorithm \ref{alg.optweights} and $v_i=1-\beta_i$. The proposed near-optimal weights are obtained using support distribution $\{\beta_k\}_{k=1}^p$ and pdf of the coefficients i.e. $\{f_{d_k}(\cdot)\}_{k=1}^p$ as prior information. Figure \ref{fig.recovery_error} shows the error mean defined by $\mathds{E}_{x,\bm{A}}\|\bm{x}-\widehat{\bm{x}}\|_2$ where $\widehat{\bm{x}}$ is the obtained estimate by solving $\mathsf{P}_{\bm{\Omega},\bm{v}}$. The latter expectation is calculated using $100$ independent Monte Carlo simulations over different $\bm{A}$ and $\bm{x}$. 
 As shown in Figure \ref{fig.recovery_error}, our near-optimal weights provides a superior performance than the heuristic and constant weights in recovering the interested signal from under-determined measurements $\bm{y}\in\mathbb{R}^m$.
 
\begin{figure}[t]
	\hspace{0cm}
	\includegraphics[scale=.4]{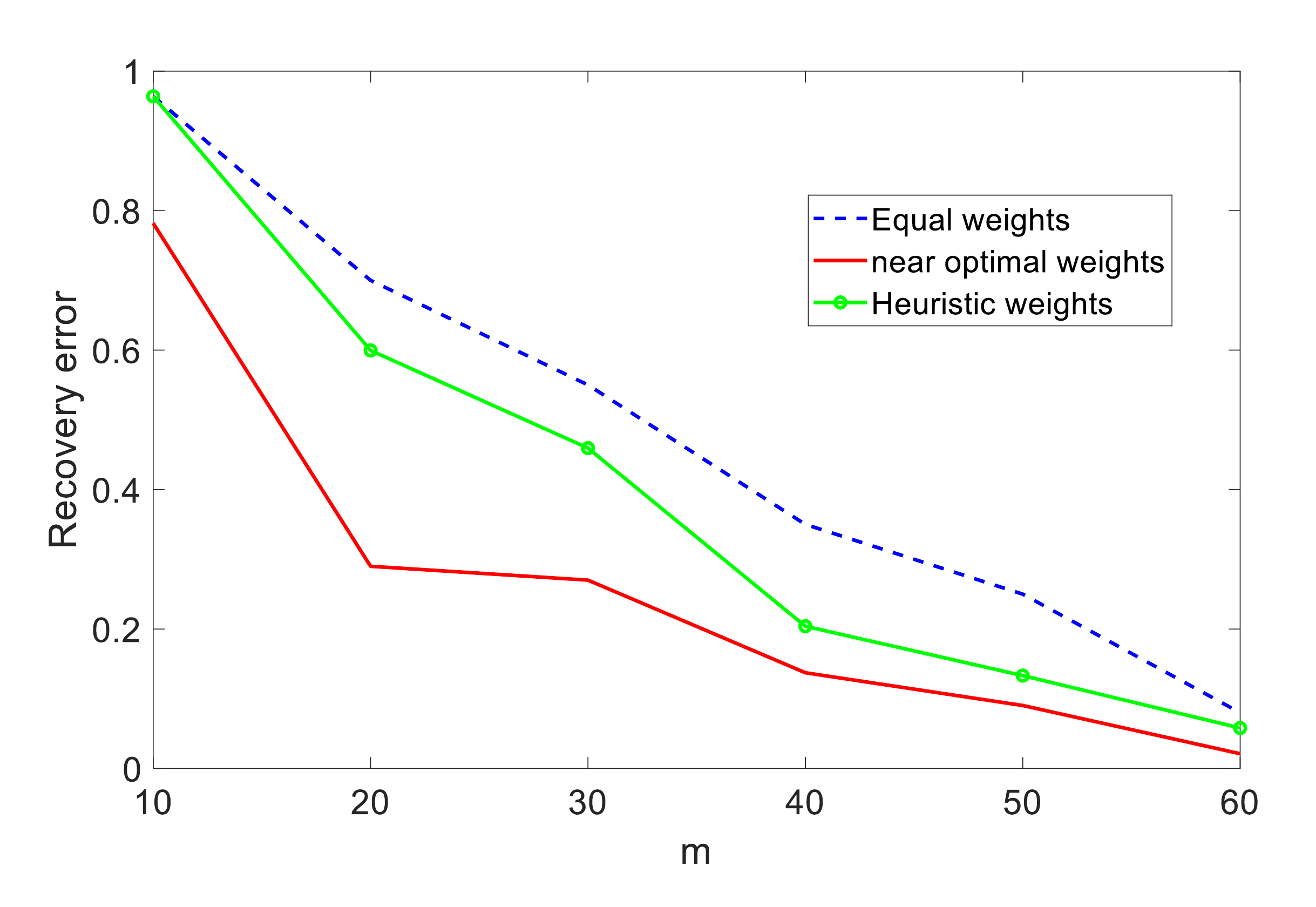}
	\caption{Recovery performance of the MRI experiment in Section \ref{sec.simulation}. We examined three different cases of weights: Heuristic, constant and near-optimal in problem $\mathsf{P}_{\bm{\Omega},\bm{v}}$ \eqref{eq.weightedl1}. The heuristic weights are selected as $v_i=1-\beta_i, i=1,..., p$ and the near-optimal weights are found by Algorithm \ref{alg.optweights}.}\label{fig.recovery_mri}
\end{figure}

 In the second experiment, we use an MRI sequence which is composed of $27$ frames of size $128 \times 128$. We first resized the images to $64\times 64$ to be able to process them computationally. We employed a two-dimensional wavelet analysis operator as $\bm{\Omega}$ with $p=256, n=64$ of daubechies family. The number of decomposition levels is fixed to $1$ while the filter length is set to $8$. To generate this operator, we used SPOT package \cite{van2014spot}. We selected the $27$-th image and used the other $26$ images to obtain the empirical distribution which are employed to find the near-optimal weights according to Theorem \ref{thm.main}. We repeated the experiment for all of the images and obtain the expected $\mathds{E}_{\bm{x},\bm{A}}\|\bm{x}-\widehat{\bm{x}}\|_2$. We examined the performance of our method compared to $\mathsf{P}_{\bm{\Omega},\bm{v}}$ with heuristic ($v_i=1-\beta_i, i=1,..., p$) and constant weights ($v_i=1, i=1,...,p$) in  Figure \ref{fig.recovery_mri}. As it turns out, weighted $\ell_1$ minimization with near-optimal weights needs fewer measurements compared to heuristic approach to reach a same level of performance. 
 \begin{figure}[t]
 	\centering
 	\includegraphics[scale=.5]{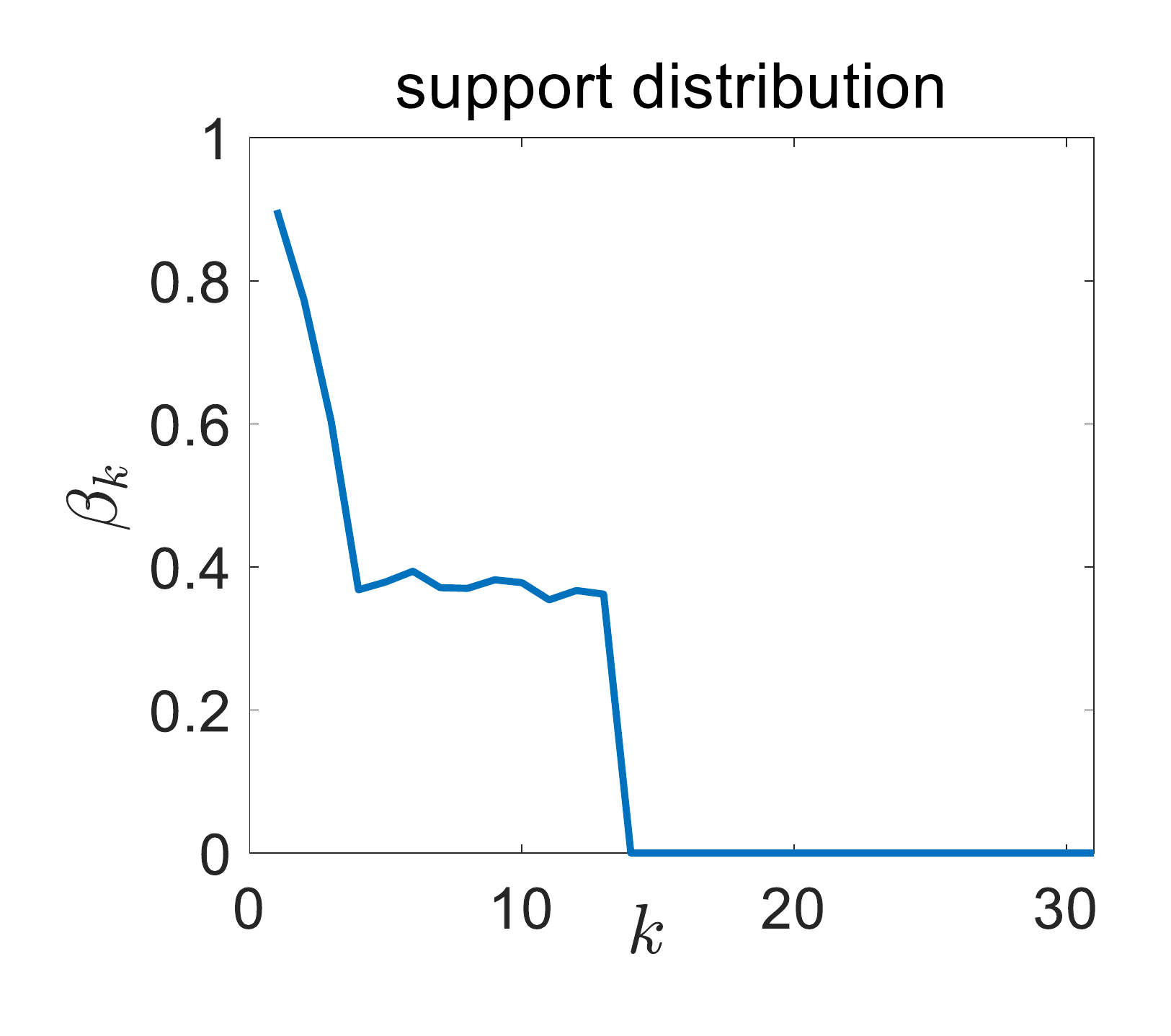}
 	\caption{The support distribution in the first experiment of Section \ref{sec.simulation}. Based on this distribution and the expected sign values, the near optimal weights are determined  using Algorithm \ref{alg.optweights}.}\label{fig.support_dist}
 \end{figure}
 \begin{figure}[t]
	\hspace{.3cm}
	\includegraphics[scale=.35]{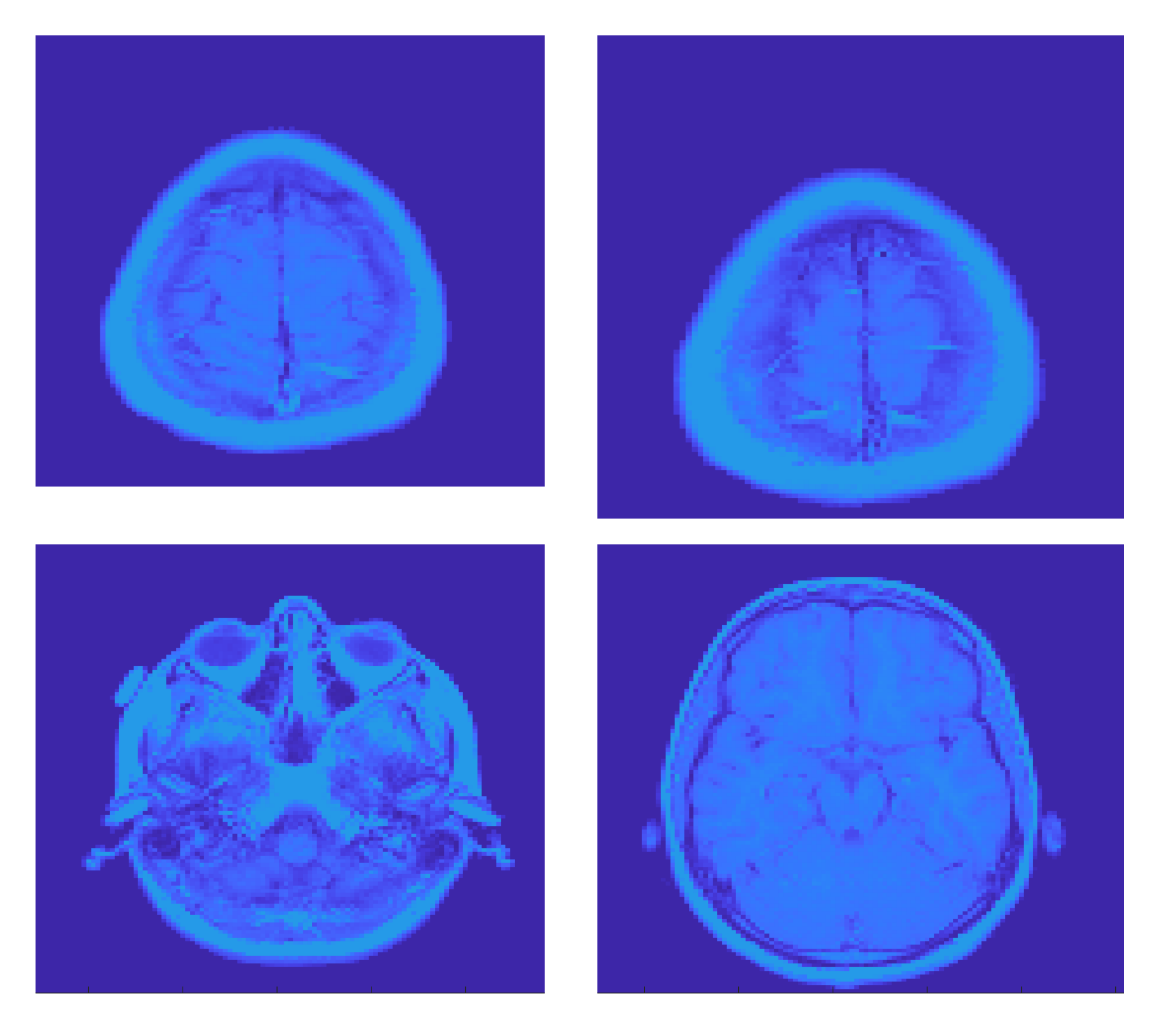}
	\caption{A few sample frames of MRI scan used in Section \ref{sec.simulation}. Top right: $27$-th frame. Top left: $26$-th frame. Bottom left: $3$-th frame. Bottom right: $10$-th frame. The performance of sampling  $27$-th frame is evaluated using prior distribution obtained from previous $26$ MRI frames.}\label{fig.mri}
\end{figure}

 \section{Conclusion}\label{sec.conclusion}
 In this paper, we developed a novel framework to incorporate statistical information into the recovery of analysis-sparse signals. First, we proposed a tight upper-bound describing the expected number of measurements that weighted $\ell_1$ analysis minimization needs for exact recovery. The bound depends on support distribution and expected sign of coefficients which are both achievable in advance. Then, we obtained the near-optimal weights by minimizing this bound with respect to the weights. As simulation results verify, our near-optimal weights lead to an enhanced performance and less error reconstruction compared to regular $\ell_1$ analysis minimization.
 
\appendix
\subsection{Proof of Lemma \ref{lem.upper1}}\label{proof.lemma_upper}
First, we begin with the definition of statistical dimension, i.e.
\begin{align}\label{eq.rel1}
&\delta(\mathcal{D}(\|\bm{\Omega}\cdot\|_{1,\bm{v}},\bm{x}))= \mathds{E}_{\bm{g}}{\rm dist}^2(\bm{g},{\rm cone}\partial \|\bm{\Omega}\cdot\|_1(\bm{x}))=\nonumber\\
&\mathds{E}_{\bm{g}} \inf_{t\ge 0}\inf_{\bm{z}\in\partial \|\bm{\Omega} \cdot\|_{1,\bm{v}}({\bm{x}})}\|\bm{g}-t\bm{z}\|_2^2.
\end{align}
By passing $\mathds{E}_{\bm{g}}$ through the first infimum, we reach an upper-bound as follows:
\begin{align}\label{eq.rel2}
&\mathds{E}_{\bm{g}}\inf_{t\ge 0}\inf_{\bm{z}\in\partial \|\bm{\Omega} \cdot\|_1({\bm{x}})}\|\bm{g}-t\bm{z}\|_2^2\le \inf_{t\ge 0}\mathds{E}_{\bm{g}}
\inf_{\bm{z}\in\partial \|\bm{\Omega} \cdot\|_{1,\bm{v}}({\bm{x}})}\|\bm{g}-t\bm{z}\|_2^2\nonumber\\
&=\inf_{t\ge 0}\mathds{E}_{\bm{g}}
\inf_{\|\bm{z}\|_{\infty}\le 1}\|\bm{g}-t\bm{\Omega}^T\big(\bm{v}\odot{\rm sgn}(\bm{\Omega x})\big)-t\bm{\Omega}^T\big(\bm{v}\odot\bm{z}\big)\|_2^2\nonumber\\
&:=B_u,
\end{align}
where in the last expression above, we used the chain rule of subdifferential (see \cite[Chapter 23]{rockafellar1970convex}) i.e. ${\partial\|\bm{\Omega}\cdot\|_{1,\bm{v}}=\bm{\Omega}^T\partial \|\cdot\|_{1,\bm{v}}(\bm{\Omega x})}$ and the well-established relation for subdifferential of norm functions (see for example \cite[Proposition 1]{daei2019exploiting} or \cite[Proposition 6]{flinth2016optimal}) i.e.
\begin{align}
\partial \|\cdot\|_{1,\bm{v}}(\bm{\Omega x})=\{\bm{v}\odot\big({\rm sgn}(\bm{\Omega x})+\bm{z}\big): ~\|\bm{z}\|_{\infty}\le 1\}.
\end{align}
In \cite{daei2019error}, it is shown that $B_u$ is tight for different kinds of analysis operators. The expression $B_u$ does not seem to have a closed-form formula in general. Hence, we seek for an upper-bound of $B_u$. For this purpose, we substitute $\bm{z}$ with a special $\bm{z}'=[z_1',..., z_p']^p$ whose elements are defined as 
\begin{align}\label{eq.choice}
{z}_i':=\left\{\begin{array}{lr}
0,&i\in\mathcal{S}\\
{\rm sgn}(\bm{\omega}_i^T\bm{g})\big(\frac{(tv_i)^{-1}\lambda |\bm{\omega}_i^T\bm{g}|}{\|\bm{\omega}_i\|_2}\wedge 1\big),&i\in\overline{\mathcal{S}}
\end{array}\right\},
\end{align}
where $\lambda>0$ is a flexible tuning parameter. With this choice of $\bm{z}$, we can proceed with \eqref{eq.rel2} as follows:
\begin{align}\label{eq.rel4}
&\mathds{E}_{\bm{g}}
\|\bm{g}-t\bm{\Omega}^T\big(\bm{v}\odot{\rm sgn}(\bm{\Omega x})\big)-t\bm{\Omega}^T\big(\bm{z}'\odot \bm{v}\big)\|_2^2=\nonumber\\
&\mathds{E}_{\bm{g}}\|\bm{g}\|_2^2-
2t\underbrace{\mathds{E}_{\bm{g}}\langle \bm{g},\bm{\Omega}^T\big(\bm{v}\odot{\rm sgn}(\bm{\Omega x})\big)\rangle}_{0}\nonumber\\
&-\underbrace{2t\mathds{E}_{\bm{g}}\langle \bm{g}, \bm{\Omega}^T\big(\bm{v}\odot\bm{z}'\big)\rangle}_{:=V_1}+\underbrace{t^2\mathds{E}_{\bm{g}}\|\bm{\Omega}^T\big(\bm{v}\odot\bm{z}'\big)\|_2^2}_{:=V_2}+\nonumber\\
&\underbrace{t^2\|\bm{\Omega}^T\big(\bm{v}\odot{\rm sgn}(\bm{\Omega x})\big)\|_2^2}_{:=V_3}=n-V_1+V_2+V_3.
\end{align}
Now, we calculate each term above one by one. First, consider
\begin{align}
&V_1=2t\mathds{E}_{\bm{g}}\langle \bm{\Omega}\bm{g},\bm{v}\odot\bm{z}'\rangle=2t\sum_{i\in\overline{\mathcal{S}}}v_i\mathds{E}_{\bm g}\big[\bm{\omega}_i^T\bm{g}z_i'\big]\nonumber\\
&
=2\sum_{i\in\overline{\mathcal{S}}}\mathds{E}_{\bm g}\bigg[|\bm{\omega}_i^T\bm{g}|\big(\frac{\lambda |\bm{\omega}_i^T\bm{g}|}{\|\bm{\omega}_i\|_2}\wedge tv_i\big)\bigg],
\end{align}
where we used the relation \eqref{eq.choice} in the last expression. Since $\bm{\omega}_i^T\bm{g}\sim \mathcal{N}(\bm{0},\|\bm{\omega}_i\|_2^2)$, it is straightforward to obtain the expectation inside the summation and verify that
\begin{align}
V_1=2\sum_{i\in\overline{\mathcal{S}}}\|\bm{\omega}_i\|_2\mathds{E}_{\bm{g}}\big[|\bm{g}|(\lambda |\bm{g}|\wedge tv_i)\big].
\end{align}
We proceed by calculating
\begin{align}
&\mathds{E}_{\bm{g}}\big[|\bm{g}|(\lambda |\bm{g}|\wedge tv_i)\big]=\int_{0}^{\frac{tv_i}{\lambda}}\lambda\sqrt{\frac{2}{\pi}}z^2 {\rm e}^{-\frac{z^2}{2}}{\rm d}z+\nonumber\\
&\int_{\frac{tv_i}{\lambda}}^{\infty}\sqrt{\frac{2}{\pi}}t v_i z {\rm e}^{-\frac{z^2}{2}}{\rm d}z=\sqrt{\frac{2}{\pi}} t v_i {\rm e}^{-\frac{t^2v_i^2}{2\lambda^2}}-\sqrt{\frac{2}{\pi}} t v_i {\rm e}^{-\frac{t^2v_i^2}{2\lambda^2}}+\nonumber\\
&\lambda {\rm erf}(\frac{tv_i}{\sqrt{2}\lambda}),
\end{align}
which leads to
\begin{align}\label{eq.V1relfinal}
V_1=2\sum_{i\in\overline{\mathcal{S}}}\lambda \|\bm{\omega}_i\|_2 {\rm erf}(\frac{tv_i}{\sqrt{2}\lambda}).
\end{align}
For $V_2$, we have
\begin{align}\label{eq.V2}
&V_2=t^2\mathds{E}_{\bm{g}}\langle \bm{v}\odot\bm{z}', \bm{\Omega}\bm{\Omega}^T \big(\bm{v}\odot\bm{z}'\big)\rangle=t^2\sum_{i,j\in\overline{\mathcal{S}}}v_iv_j\bm{\omega}_i^T\bm{\omega}_j\nonumber\\
&\mathds{E}_{\bm{g}}[z_i'z_j']=\sum_{i,j\in\overline{\mathcal{S}}}
(tv_i)(tv_j)\bm{\omega}_i^T\bm{\omega}_j\mathds{E}_{\bm{g}}\bigg[{\rm sgn}(\bm{\omega}_i^T\bm{g}){\rm sgn}(\bm{\omega}_j^T\bm{g})\nonumber\\
&\big(\frac{(tv_i)^{-1}\lambda \bm{\omega}_i^T\bm{g}}{\|\bm{\omega}_i\|_2}\wedge 1\big)\big(\frac{(tv_j)^{-1}\lambda \bm{\omega}_j^T\bm{g}}{\|\bm{\omega}_j\|_2}\wedge 1\big)\bigg]=\sum_{i,j\in\overline{\mathcal{S}}}\lambda^2\bm{\omega}_i^T\bm{\omega}_j\nonumber\\
&\mathds{E}_{\bm{g}}\bigg[{\rm sgn}(\bm{\omega}_i^T\bm{g}){\rm sgn}(\bm{\omega}_j^T\bm{g})\big(\frac{|\bm{\omega}_i^T\bm{g}|}{\|\bm{\omega}_i\|_2}\wedge \frac{tv_i}{\lambda}\big)\big(\frac{|\bm{\omega}_j^T\bm{g}|}{\|\bm{\omega}_j\|_2}\wedge \frac{tv_j}{\lambda}\big)\bigg]
\end{align}

Obtaining a closed-form formula for the last expression above seems to be difficult in general. For this case, we borrow an upper-bound in the following Lemma which is borrowed from \cite[Lemma 6.12]{genzel2017ell}.
\begin{lem}\cite[Lemma 6.12]{genzel2017ell}
Let $\bm{g}\sim \mathcal{N}(\bm{0}, \bm{I}_n)$. For any $\bm{z}_1, \bm{z}_2\in\mathbb{S}^{n-1}$, $\alpha_1, \alpha_2\ge 0$, and $\alpha_{\min}=\min\{\alpha_1,\alpha_2\}$, $\alpha_{\max}=\max\{\alpha_1,\alpha_2\}$, we have
\begin{align}
&\mathds{E}\bigg[{\rm sgn}(\bm{z}_1^T\bm{g}){\rm sgn}(\bm{z}_2^T\bm{g})\big(|\bm{z}_1^T\bm{g}|\wedge \alpha_1\big)\big(|\bm{z}_2^T\bm{g}|\wedge \alpha_2\big)\bigg]\le \nonumber\\
&|\langle \bm{z}_1, \bm{z}_2 \rangle|\bigg[ {\rm erf}(\frac{\alpha_{\min}}{\sqrt{2}})-h(\alpha_{\max})\alpha_1\alpha_2\bigg].
\end{align}
\end{lem}
By benefiting this lemma, we proceed \eqref{eq.V2} as follows:
 \begin{align}\label{eq.V2relfinal}
 V_2\le \lambda^2\sum_{i,j\in\overline{\mathcal{S}}}\frac{(\bm{\omega}_i^T\bm{\omega}_j)^2}{\|\bm{\omega}_i\|_2\|\bm{\omega}_j\|_2}\bigg[{\rm erf}(\frac{tv_{\min}}{\lambda\sqrt{2}})-h(\frac{tv_{\max}}{\lambda})\frac{(tv_1)(tv_2)}{\lambda^2}\bigg].
 \end{align}
 Lastly, for $V_3$, it holds that
 \begin{align}\label{eq.V3relfinal}
 &V_3=t^2\langle \bm{v}\odot {\rm sgn}(\bm{\Omega x}), \bm{\Omega}\bm{\Omega}^T \big(\bm{v}\odot {\rm sgn}(\bm{\Omega x})\big)\rangle=\nonumber\\
 &\sum_{i,j\in\mathcal{S}}(tv_i)(tv_j)\bm{\omega}_i^T\bm{\omega}_j{\rm sgn}(\bm{\omega}_i^T\bm{x}){\rm sgn}(\bm{\omega}_j^T\bm{x}).
 \end{align}
 Combining \eqref{eq.V1relfinal}, \eqref{eq.V2relfinal}, \eqref{eq.V3relfinal}, and \eqref{eq.rel4}, leads to the following upper-bound for $\delta(\mathcal{D}(\|\bm{\Omega}\cdot\|_{1,\bm{v}},\bm{x}))$.
 \begin{align}
 &\delta(\mathcal{D}(\|\bm{\Omega}\cdot\|_{1,\bm{v}},\bm{x}))\le \inf_{t> 0}\inf_{\lambda> 0}\Bigg\{n+\sum_{i,j\in\mathcal{S}}(tv_i)(tv_j)\bm{\omega}_i^T\bm{\omega}_j\nonumber\\
 &{\rm sgn}(\bm{\omega}_i^T\bm{x}){\rm sgn}(\bm{\omega}_j^T\bm{x})-2\sum_{i\in\overline{\mathcal{S}}}\lambda \|\bm{\omega}_i\|_2 {\rm erf}(\frac{tv_i}{\sqrt{2}\lambda})+\nonumber\\
 &\lambda^2\sum_{i,j\in\overline{\mathcal{S}}}\frac{(\bm{\omega}_i^T\bm{\omega}_j)^2}{\|\bm{\omega}_i\|_2\|\bm{\omega}_j\|_2}\bigg[{\rm erf}(\frac{tv_{\min}}{\lambda\sqrt{2}})-h(\frac{tv_{\max}}{\lambda})\frac{(tv_1)(tv_2)}{\lambda^2}\bigg]\Bigg\}.
 \end{align}
 By substituting $t$ by $\lambda t$, we have
\begin{align}
&\delta(\mathcal{D}(\|\bm{\Omega}\cdot\|_{1,\bm{v}},\bm{x}))\le \inf_{t> 0}\inf_{\lambda> 0}\Bigg\{n+\lambda^2\sum_{i,j\in\mathcal{S}}(tv_i)(tv_j)\bm{\omega}_i^T\bm{\omega}_j\nonumber\\
&{\rm sgn}(\bm{\omega}_i^T\bm{x}){\rm sgn}(\bm{\omega}_j^T\bm{x})-2\sum_{i\in\overline{\mathcal{S}}}\lambda \|\bm{\omega}_i\|_2 {\rm erf}(\frac{tv_i}{\sqrt{2}})+\nonumber\\
&\lambda^2\sum_{i,j\in\overline{\mathcal{S}}}\frac{(\bm{\omega}_i^T\bm{\omega}_j)^2}{\|\bm{\omega}_i\|_2\|\bm{\omega}_j\|_2}\bigg[{\rm erf}(\frac{tv_{\min}}{\sqrt{2}})-h(tv_{\max})(tv_1)(tv_2)\bigg]\Bigg\}.
\end{align}
\subsection{Proof of Theorem \ref{thm.main}}\label{proof.thm_main}
By incorporating the randomness of $\bm{x}$ and thus $\bm{d}:=\bm{\Omega x}$, we obtain an upper-bound for $\mathds{E}_{\bm{x}}\delta(\mathcal{D}(\|\bm{\Omega} \cdot\|_{1,\bm{v}},\bm{x}))$ as follows. First, by taking expectation with respect to $\bm{x}$ from \eqref{eq.upper1} in Lemma \ref{lem.upper1}, it follows that
\begin{align}
&\mathds{E}_{\bm{x}}\delta(\mathcal{D}(\|\bm{\Omega} \cdot\|_{1,\bm{v}},\bm{x}))\le \inf_{t> 0}\inf_{\lambda> 0}\Bigg\{n+\lambda^2\sum_{i,j=1}^p (tv_i)(tv_j)\nonumber\\
&\bm{\omega}_i^T\bm{\omega}_j\mathds{E}_{\bm{x}}\big[{\rm sgn}(\bm{\omega}_i^T\bm{x}){\rm sgn}(\bm{\omega}_j^T\bm{x})\big]\nonumber\\
&-2\sum_{i=1}^p\lambda \|\bm{\omega}_i\|_2 {\rm erf}(\frac{tv_i}{\sqrt{2}})\mathds{E}_{\bm{x}}\big[1_{i\in\overline{\mathcal{S}}}\big]+\lambda^2\sum_{i,j=1}^p\frac{(\bm{\omega}_i^T\bm{\omega}_j)^2}{\|\bm{\omega}_i\|_2\|\bm{\omega}_j\|_2}\nonumber\\
&\bigg[{\rm erf}(\frac{tv_{\min}}{\sqrt{2}})-h(tv_{\max})(tv_i)(tv_j)\bigg]\mathds{E}_{\bm{x}}\big[1_{i\in\overline{\mathcal{S}}}1_{j\in\overline{\mathcal{S}}}\big]\Bigg\}.
\end{align}
By further simplifying, we can reach
\begin{align}
&\mathds{E}_{\bm{x}}\delta(\mathcal{D}(\|\bm{\Omega} \cdot\|_{1,\bm{v}},\bm{x}))\le \inf_{t> 0}\inf_{\lambda> 0}\nonumber\\
&\Bigg\{n+\lambda^2\sum_{i=1}^p (tv_i)^2\|\bm{\omega}_i\|_2^2\mathds{E}\big[1_{i\in\mathcal{S}}\big]+\lambda^2\sum_{\substack{i,j=1\\i\neq j}}^p(tv_i)(tv_j)\bm{\omega}_i^T\bm{\omega}_j\nonumber\\
&\mathds{E}_{\bm{x}}\big[{\rm sgn}(\bm{\omega}_i^T\bm{x}){\rm sgn}(\bm{\omega}_j^T\bm{x})\big]-2\sum_{i=1}^p\lambda \|\bm{\omega}_i\|_2 {\rm erf}(\frac{tv_i}{\sqrt{2}})\mathds{E}_{\bm{x}}\big[1_{i\in\overline{\mathcal{S}}}\big]\nonumber\\
&+\lambda^2\sum_{i=1}^p\|\bm{\omega}_i\|_2^2\bigg[{\rm erf}(\frac{tv_i}{\sqrt{2}})-h(tv_i)(tv_i)^2\bigg]\mathds{E}_{\bm{x}}\big[1_{i\in\overline{\mathcal{S}}}\big]+
\nonumber\\
&\lambda^2\sum_{\substack{i,j=1\\i\neq j}}^p\frac{(\bm{\omega}_i^T\bm{\omega}_j)^2}{\|\bm{\omega}_i\|_2\|\bm{\omega}_j\|_2}\bigg[{\rm erf}(\frac{tv_{\min}}{\sqrt{2}})-h(tv_{\max})(tv_i)(tv_j)\bigg]\nonumber\\
&\mathds{E}_{\bm{x}}\big[1_{i\in\overline{\mathcal{S}}}1_{j\in\overline{\mathcal{S}}}\big]\Bigg\}.
\end{align}
By using the definitions in Theorem \ref{thm.main}, we have
\begin{align}\label{eq.upp1}
&\mathds{E}_{\bm{x}}\delta(\mathcal{D}(\|\bm{\Omega} \cdot\|_{1,\bm{v}},\bm{x}))\le \inf_{t> 0}\inf_{\lambda> 0}\nonumber\\
&\Bigg\{n+\lambda^2\sum_{i=1}^p (tv_i)^2\|\bm{\omega}_i\|_2^2\beta_i+\lambda^2\sum_{\substack{i,j=1\\i\neq j}}^p(tv_i)(tv_j)\bm{\omega}_i^T\bm{\omega}_j\nonumber\\
&\sigma_{ij}\beta_i\beta_j-2\sum_{i=1}^p\lambda \|\bm{\omega}_i\|_2 {\rm erf}(\frac{tv_i}{\sqrt{2}})(1-\beta_i)+\lambda^2\sum_{i=1}^p\|\bm{\omega}_i\|_2^2\nonumber\\
&\bigg[{\rm erf}(\frac{tv_i}{\sqrt{2}})-h(tv_i)(tv_i)^2\bigg](1-\beta_i)+
\lambda^2\sum_{\substack{i,j=1\\i\neq j}}^p\frac{(\bm{\omega}_i^T\bm{\omega}_j)^2}{\|\bm{\omega}_i\|_2\|\bm{\omega}_j\|_2}\nonumber\\
&\bigg[{\rm erf}(\frac{tv_{\min}}{\sqrt{2}})-h(tv_{\max})(tv_i)(tv_j)\bigg](1-\beta_i)(1-\beta_j)\Bigg\}.
\end{align}
 According to Figure \ref{fig.lem1}, the latter bound is numerically observed to be strictly convex with respect to $(t,\lambda)$. By taking the derivative of the latter relation with respect to $\lambda$ and setting it to zero, we have:
 \begin{align}
 &2\lambda\Bigg\{\sum_{i=1}^p (tv_i)^2\|\bm{\omega}_i\|_2^2\beta_i+\sum_{\substack{i,j=1\\i\neq j}}^p(tv_i)(tv_j)\bm{\omega}_i^T\bm{\omega}_j\sigma_{ij}\beta_i\beta_j+\nonumber\\
 &\sum_{i=1}^p\|\bm{\omega}_i\|_2^2\bigg[{\rm erf}(\frac{tv_i}{\sqrt{2}})-h(tv_i)(tv_i)^2\bigg](1-\beta_i)+
 \sum_{\substack{i,j=1\\i\neq j}}^p\frac{(\bm{\omega}_i^T\bm{\omega}_j)^2}{\|\bm{\omega}_i\|_2\|\bm{\omega}_j\|_2}\nonumber\\
 &\bigg[{\rm erf}(\frac{tv_{\min}}{\sqrt{2}})-h(tv_{\max})(tv_i)(tv_j)\bigg](1-\beta_i)(1-\beta_j) \Bigg\}\nonumber\\
 &-2\sum_{i=1}^p \|\bm{\omega}_i\|_2 {\rm erf}(\frac{tv_i}{\sqrt{2}})(1-\beta_i)=0
 \end{align}
 which leads to
 \begin{align}
 \lambda^\star=\frac{\sum_{i=1}^p \|\bm{\omega}_i\|_2 {\rm erf}(\frac{tv_i}{\sqrt{2}})(1-\beta_i)}
 {F(t,v)}.
 \end{align}
 By replacing this $\lambda$ into \eqref{eq.upp1}, the result of Theorem \ref{thm.main} is achieved.
\ifCLASSOPTIONcaptionsoff
  \newpage
\fi

\bibliographystyle{ieeetr}
\bibliography{mypaperbibe}

\begin{thebibliography}{10}

\bibitem{candes2008restricted}
E.~J. Candes, ``The restricted isometry property and its implications for
  compressed sensing,'' {\em Comptes rendus mathematique}, vol.~346, no.~9-10,
  pp.~589--592, 2008.

\bibitem{candes2005decoding}
E.~J. Candes and T.~Tao, ``Decoding by linear programming,'' {\em IEEE
  transactions on information theory}, vol.~51, no.~12, pp.~4203--4215, 2005.

\bibitem{donoho2005sparse}
D.~L. Donoho and J.~Tanner, ``Sparse nonnegative solution of underdetermined
  linear equations by linear programming,'' {\em Proceedings of the National
  Academy of Sciences of the United States of America}, vol.~102, no.~27,
  pp.~9446--9451, 2005.

\bibitem{donoho2006high}
D.~L. Donoho, ``High-dimensional centrally symmetric polytopes with
  neighborliness proportional to dimension,'' {\em Discrete \& Computational
  Geometry}, vol.~35, no.~4, pp.~617--652, 2006.

\bibitem{candes2011compressed}
E.~J. Candes, Y.~C. Eldar, D.~Needell, and P.~Randall, ``Compressed sensing
  with coherent and redundant dictionaries,'' {\em Applied and Computational
  Harmonic Analysis}, vol.~31, no.~1, pp.~59--73, 2011.

\bibitem{genzel2017ell}
M.~Genzel, G.~Kutyniok, and M.~März, ``$\ell_1$-analysis minimization and
  generalized (co-)sparsity: When does recovery succeed?,'' {\em Applied and
  Computational Harmonic Analysis}, vol.~52, pp.~82--140, 2021.

\bibitem{daei2019error}
S.~Daei, F.~Haddadi, A.~Amini, and M.~Lotz, ``On the error in phase transition
  computations for compressed sensing,'' {\em IEEE Transactions on Information
  Theory}, vol.~65, no.~10, pp.~6620--6632, 2019.

\bibitem{daei2019living}
S.~Daei, F.~Haddadi, and A.~Amini, ``Living near the edge: A lower-bound on the
  phase transition of total variation minimization,'' {\em IEEE Transactions on
  Information Theory}, vol.~66, no.~5, pp.~3261--3267, 2019.

\bibitem{candes2006robust}
E.~J. Cand{\`e}s, J.~Romberg, and T.~Tao, ``Robust uncertainty principles:
  Exact signal reconstruction from highly incomplete frequency information,''
  {\em IEEE Transactions on information theory}, vol.~52, no.~2, pp.~489--509,
  2006.

\bibitem{amelunxen2013living}
D.~Amelunxen, M.~Lotz, M.~B. McCoy, and J.~A. Tropp, ``Living on the edge:
  Phase transitions in convex programs with random data,'' {\em Information and
  Inference: A Journal of the IMA}, vol.~3, no.~3, pp.~224--294, 2014.

\bibitem{daei2018sample}
S.~Daei, F.~Haddadi, and A.~Amini, ``Sample complexity of total variation
  minimization,'' {\em IEEE Signal Processing Letters}, vol.~25, no.~8,
  pp.~1151--1155, 2018.

\bibitem{daei2019exploiting}
S.~Daei, F.~Haddadi, and A.~Amini, ``Exploiting prior information in
  block-sparse signals,'' {\em IEEE Transactions on Signal Processing},
  vol.~67, no.~19, pp.~5093--5102, 2019.

\bibitem{daei2018improved}
S.~Daei, F.~Haddadi, and A.~Amini, ``Improved recovery of analysis sparse
  vectors in presence of prior information,'' {\em IEEE Signal Processing
  Letters}, vol.~26, no.~2, pp.~222--226, 2018.

\bibitem{daei2019distribution}
S.~Daei, F.~Haddadi, and A.~Amini, ``Distribution-aware block-sparse recovery
  via convex optimization,'' {\em IEEE Signal Processing Letters}, vol.~26,
  no.~4, pp.~528--532, 2019.

\bibitem{flinth2016optimal}
A.~Flinth, ``Optimal choice of weights for sparse recovery with prior
  information,'' {\em IEEE Transactions on Information Theory}, vol.~62, no.~7,
  pp.~4276--4284, 2016.

\bibitem{diaz2017compressed}
M.~D{\'\i}az, M.~Junca, F.~Rinc{\'o}n, and M.~Velasco, ``Compressed sensing of
  data with a known distribution,'' {\em Applied and Computational Harmonic
  Analysis}, 2017.

\bibitem{rockafellar2015convex}
R.~T. Rockafellar, {\em Convex analysis}.
\newblock Princeton university press, 2015.

\bibitem{chandrasekaran2012convex}
V.~Chandrasekaran, B.~Recht, P.~A. Parrilo, and A.~S. Willsky, ``The convex
  geometry of linear inverse problems,'' {\em Foundations of Computational
  mathematics}, vol.~12, no.~6, pp.~805--849, 2012.

\bibitem{daei2018optimal}
S.~Daei, A.~Amini, and F.~Haddadi, ``Optimal weighted low-rank matrix recovery
  with subspace prior information,'' {\em arXiv preprint arXiv:1809.10356},
  2018.

\bibitem{cvx}
M.~Grant and S.~Boyd, ``Cvx: Matlab software for disciplined convex
  programming, version 2.1,'' 2014.

\bibitem{kabanava2015analysis}
M.~Kabanava and H.~Rauhut, ``Analysis $\ell_1$-recovery with frames and
  {G}aussian measurements,'' {\em Acta Applicandae Mathematicae}, vol.~140,
  no.~1, pp.~173--195, 2015.

\bibitem{nam2013cosparse}
S.~Nam, M.~E. Davies, M.~Elad, and R.~Gribonval, ``The cosparse analysis model
  and algorithms,'' {\em Applied and Computational Harmonic Analysis}, vol.~34,
  no.~1, pp.~30--56, 2013.

\bibitem{van2014spot}
E.~Van~den Berg and M.~Friedlander, ``Spot-a linear-operator toolbox,'' {\em
  URL http://www. cs. ubc. ca/labs/scl/spot}, 2014.

\bibitem{rockafellar1970convex}
R.~T. Rockafellar, ``Convex analysis princeton university press,'' {\em
  Princeton, NJ}, 1970.

\end{thebibliography}
\end{document}